\documentclass[showpacs,prl,onecolumn,aps,superscriptaddress,preprintnumbers,nofootinbib]{revtex4}
\usepackage[T1]{fontenc}
\usepackage[latin9]{inputenc}
\usepackage{amsmath,amssymb}
\usepackage{epsfig}
\usepackage{graphicx}
\usepackage{amsmath}
\usepackage{amsfonts}
\def\slashchar#1{\setbox0=\hbox{$#1$}     		
   \dimen0=\wd0                                 	
   \setbox1=\hbox{/} \dimen1=\wd1               	
   \ifdim\dimen0>\dimen1                        	
      \rlap{\hbox to \dimen0{\hfil/\hfil}}      	
      #1                                        	
   \else                                        	
      \rlap{\hbox to \dimen1{\hfil$#1$\hfil}}   	
      /                                         	
   \fi}

\renewcommand{\vec}{\boldsymbol}
\newcommand{\beq}{\begin{equation}}
\newcommand{\eeq}{\end{equation}}
\newcommand{\bea}{\begin{eqnarray}}
\newcommand{\eea}{\end{eqnarray}}
\newcommand{\baa}{\begin{array}}
\newcommand{\eaa}{\end{array}}

\def\eq#1{{Eq.~(\ref{#1})}}

\def\fig#1{{Fig.~\ref{#1}}}

\newcommand{\bas}{\bar{\alpha}_S}
\newcommand{\as}{\alpha_S}

\newcommand{\nn}{\nonumber}

\newcommand{\h}{\frac{1}{2}}

\newcommand{\Lb}{\left(}
\newcommand{\Rb}{\right)}

\renewcommand{\vec}[1]{\boldsymbol{#1}}

\begin{document}
\title{BFKL equation in the next-to-leading order: solution at large impact parameters.}
\author{Carlos Contreras}
\email{carlos.contreras@usm.cl}
\affiliation{Departamento de F\'isica, Universidad T\'ecnica Federico Santa Mar\'ia, and Centro Cient\'ifico-\\
Tecnol\'ogico de Valpara\'iso, Avda. Espa\~na 1680, Casilla 110-V, Valpara\'iso, Chile}
\author{ Eugene ~ Levin}
\email{leving@tauex.tau.ac.il, eugeny.levin@usm.cl}
\affiliation{Departamento de F\'isica, Universidad T\'ecnica Federico Santa Mar\'ia, and Centro Cient\'ifico-\\
Tecnol\'ogico de Valpara\'iso, Avda. Espa\~na 1680, Casilla 110-V, Valpara\'iso, Chile}
\affiliation{Department of Particle Physics, School of Physics and Astronomy,
Raymond and Beverly Sackler
 Faculty of Exact Science, Tel Aviv University, Tel Aviv, 69978, Israel}
\author{Rodrigo Meneses}
\email{rodrigo.meneses@uv.cl}
\affiliation{Escuela de Ingenier\'\i a Civil, Facultad de Ingenier\'\i a, Universidad de Valpara\'\i so, General Cruz 222, Valpara\'\i so, Chile}

\date{\today}

\begin{abstract}
    In this paper, we show (i)  that the NLO corrections do not change 
the power-like decrease of the scattering amplitude  at large impact 
parameter ($b^2 \,>\,r^2 \exp\Lb 2\bas \eta(1 + 4 \bas)\Rb$, where
 $r$  denotes the size of scattering dipole and 
$\eta\,=\,\ln\Lb1/x_{Bj}\Rb$
 for DIS), and, therefore, they  do not   resolve the  inconsistency 
with
  unitarity; and  (ii)    
  they  lead to an oscillating behaviour of the scattering amplitude at
 large $b$, in  direct contradiction with the unitarity constraints.
   
    However, from the more practical point of view, the NLO estimates
 give a faster decrease of the scattering amplitude as a function of
 $b$,  and could be very useful for  description of the experimental data.    
    It turns out, that  in a limited range of $b$, the NLO corrections
 generates
 the fast decrease  of the scattering amplitude with $b$, which can be
 parameterized as $N\, \propto\,\exp\Lb -\,\mu\,b\Rb$ with $\mu\,
 \propto \,1/r$ in accord with the numerical estimates in Ref.\cite{CCM}.

  \end{abstract}

\maketitle

\vspace{-0.5cm}
\tableofcontents






\section{ Introduction}

This paper is motivated and triggered by the result of the numerical
 solution\cite{CCM} of the Balitsky-Kovchegov(BK) equation in the
 next-to-leading order (NLO),  in which at large impact parameter, the
 solution shows  an  exponential decrease ($\propto \exp\Lb - \mu\, 
b\Rb$). 
 Since the amplitude decreases at large $b$, the non-linear term in  
the
 BK equation is small and can be neglected, reducing  the 
problem of large
 $b$ behaviour, to the solution of the BFKL equation.
The large impact parameter behaviour of the scattering amplitude remains
 the most  fundamental problem, which is still 
unsolved\cite{KW1,KW2,KW3} 
 in the frame of the CGC/saturation approach (see Ref.\cite{KOLEB} for a
 review). Indeed, in the CGC/saturation approach, the scattering amplitude
 decreases as a power of $b$ \cite{KW1,KW2,KW3}  contradicting
 the Froissart theorem\cite{FROI}. 
The intensive attempts to solve this problem and introduce the
 non-perturbative corrections, which  bring the dimensional
 scale into the problem\cite{LERYSL,GBS1,GKLMN,BEST1,BEST2, BEST0,LETA,LLS,LESL},
  results in the   widely held  opinion,  that we need to introduce 
a
 new non-perturbative  dimensional scale in the kernel of the BFKL equation.
 With this in mind,  the result of Ref.\cite{CCM} looks strange,
 since the NLO kernel, that has been used in the paper,  has  conformal
 symmetry, and no dimensional scale has been introduced.

The goal of this paper is to show, that in NLO we still have power-like
 behaviour at large values of $b$,  as the result of the conformal
 symmetry of the BFKL kernel. However, we find that there is a kinematic
 region where the solution has a  fast  decrease  with $b$ ($ \propto \,e^{-\mu b}$) and this 
falloff
  can be parameterized as an exponential  with $\mu\, \propto\, 1/r$, where 
$r$
 denotes the size of the scattering dipole.

The paper is organized as follows. In the next section we discuss 
 general features of the BFKL Pomeron at large values of the impact
 parameter. In Section 3, we discuss the impact parameter dependence
 in double log approximation (DLA) of the leading order of the BFKL 
 evolution equation, and show that the scattering amplitude decreases
 as a power of $b$.  Section 4 is the main part of the paper and
 it  deals with the DLA for the next-to-leading (NLA) BFKL evolution
 equation. We show that the solution for $b^2\,>\, r^2\exp\Lb \h \eta\Rb$,
 where $\eta = \ln\Lb 1/x_{Bj}\Rb$ in DIS, not only has power-like decrease
 as function of $b$, but leads to an oscillating function, which 
 contradicts   unitarity constraints. In section 5 we argue that
 the main features of the DLA  will be preserved in a more general 
approach.
 In  the  conclusion we discuss our findings, and emphasize that we need 
to
 introduce the new dimensional scale into the BFKL kernel , which is related
 to the non-perturbative corrections,  that resolve the difficulties 
at large
 $b$ in the framework of the CGC approach. On the other hand, we note that
 the NLO corrections suppress the scattering amplitude, and could  
possibly  be useful
 {\bf for} the description of the experimental data (see Ref.\cite{CCM}).

\section{  BFKL Pomeron}
 The BFKL evolution equation  for the dipole-target scattering
 amplitude $N\Lb \vec{x}_{10},\vec{b},Y \Rb$ has the general 
form\cite{KOLEB,BFKL,LIP}:
\bea \label{BFKL}
&&\frac{\partial}{\partial Y}N\Lb \vec{x}_{10},\vec{b} ,  Y \Rb = \\
&&\bas\!\! \int \frac{d^2 \vec{x}_2}{2\,\pi}\,K\Lb \vec{x}_{02}, \vec{x}_{12}; \vec{x}_{10}\Rb \Lb N\Lb \vec{x}_{12},\vec{b} - \h \vec{x}_{20}, Y\Rb + N\Lb \vec{x}_{20},\vec{b} - \h \vec{x}_{12}, \eta\Rb - N\Lb \vec{x}_{10},\vec{b},Y \Rb\Rb\nn
\eea
where $\vec{x}_{i k}\,\,=\,\,\vec{x}_i \,-\,\vec{x}_k$  and $
 \vec{x}_{10} \equiv\,\vec{r}$, $\vec{x}_{20}\,\equiv\,\vec{r}'
 $ and $\vec{x}_{12} \,\equiv\,\vec{r}\,-\,\vec{r}'$.  $Y$ is
 the rapidity of the scattering dipole and $\vec{b}$ is the
 impact factor. $K\Lb \vec{x}_{02}, \vec{x}_{12}; \vec{x}_{10}\Rb$
 is the kernel of the BFKL equation which in  leading order has
 the following form:
\beq \label{KERLO}
K_{\rm LO} \Lb \vec{x}_{02}, \vec{x}_{12}; \vec{x}_{10}\Rb\,\,=\,\,\frac{x^2_{10}}{x^2_{02}\,x^2_{12}}
\eeq

 In Ref.\cite{LIP} it has been proved that the eigenfunction of the BFKL
 equation has the following form

\beq \label{EIGENF}
\phi_\gamma\Lb \vec{r} , \vec{R}, \vec{b}\Rb\,\,\,=\,\,\,\Lb \frac{ r^2\,R^2}{\Lb \vec{b}  + \h(\vec{r} - \vec{R})\Rb^2\,\Lb \vec{b}  -  \h(\vec{r} - \vec{R})\Rb^2}\Rb^\gamma\,\,\xrightarrow{b\,\gg\,r,R}\,\,\Lb \frac{ r^2\,R^2}{b^4}\Rb^\gamma\,\,\equiv\,\,e^{\gamma\,\xi}
~~\mbox{with}~~ \xi\,=\,\ln \Lb \frac{ r^2\,R^2}{b^4}\Rb
\eeq
for any kernel,  which satisfies the conformal symmetry.
 In \eq{EIGENF} $r$ denotes the size of the scattering dipole,
 while $R$ is the size of the target. For the kernel of 
the LO BFKL equation (see \eq{KERLO}) the eigenvalues 
 take the form:

\beq \label{CHI}
\omega_{\rm LO}\Lb \bas, \gamma\Rb\,\,=\,\,\bas\,\chi^{LO}\Lb \gamma \Rb\,\,\,=\,\,\,\bas \Lb 2 \psi\Lb 1\Rb \,-\,\psi\Lb \gamma\Rb\,-\,\psi\Lb 1 - \gamma\Rb\Rb
\eeq
 $\psi(z)$  denotes the Euler psi-function $\psi\Lb z\Rb =
 d \ln \Gamma(z)/d z$.

In the next-to-leading order the kernel is derived in
 Refs.\cite{BFKLNLO,BFKLNLO1} and has the following form:
\beq \label{KERNLO1}
\omega_{\rm NLO}\Lb \bas,  \gamma\Rb\,\,=\,\,\bas\,\chi^{LO}\Lb
 \gamma \Rb\,\,+\,\,\bas^2\,\chi^{NLO}\Lb  \gamma\Rb
\eeq
The explicit form of $\chi^{NLO}\Lb  \gamma\Rb$ is given in
 Ref.\cite{BFKLNLO}. However, $\chi^{NLO}\Lb  \gamma\Rb$ turns
 out to be singular at $\gamma \to 1$,  $\chi^{NLO}\Lb\gamma\Rb
 \,\propto\,1/(1 - \gamma)^3$. Such singularities indicate,  that
  to obtain a reliable result, it is necessary 
  to calculate higher order corrections .  The procedure to re-sum 
high order corrections  is suggested
 in Ref. \cite{SALAM,SALAM1,SALAM2,KMRS}. The resulting spectrum of
 the BFKL equation in the NLO,  can be found from the solution of the
 following equation \cite{SALAM,SALAM1,SALAM2}
\beq \label{KERNLOR}
\omega_{\rm NLO}\Lb \bas,  \gamma\Rb\,=\,\bas \Lb \chi_0\Lb\omega_{\rm NLO}, \gamma\Rb\,+\,\omega_{\rm NLO} \,\frac{\chi_1\Lb \omega_{\rm NLO}, \gamma\Rb}{ \chi_0\Lb\omega_{\rm NLO}, \gamma\Rb}\Rb
\eeq
where
\beq \label{CHI0}
\chi_0\Lb\omega, \gamma\Rb\,\,=\,\,\chi^{LO}\Lb \gamma\Rb \,-\,\frac{1}{ 1 \,-\,\gamma}\,+\,\frac{1}{1\,-\,\gamma\,+\,\omega}
\eeq
and
\bea \label{CHI1}
&&\chi_1\Lb\omega, \gamma\Rb\,\,=\\
&&\,\,\chi^{NLO}\Lb \gamma\Rb\,+\,F\Lb \frac{1}{1 - \gamma}\,-\,\frac{1}{1\,-\,\gamma\,+\,\omega}\Rb\,+\,\frac{A_T\Lb \omega\Rb \,-\,A_T\Lb 0 \Rb}{\gamma^2} \,+\, \frac{A_T\Lb \omega\Rb - b}{\Lb 1\,-\,\gamma\,+\,\omega\Rb^2}\,-\,\frac{A_T\Lb 0\Rb - b}{\Lb 1\,-\,\gamma\Rb^2}\nn
 \eea
Functions $\chi^{NLO}\Lb \gamma\Rb$ and $A_T\Lb \omega\Rb$ as well as
  the constants ($F$ and $b$)  are given in 
Refs.\cite{SALAM,SALAM1,SALAM2}.

In Ref. \cite{KMRS} a  simpler  form of $\chi_1\Lb \omega,\gamma\Rb$
 was suggested,  which coincides  with \eq{CHI1} to within  $7\%$, and,
 therefore, gives reasonable estimates of all constants and functions 
in 
\eq{CHI1}.  The equation for $\omega$  takes the form
\beq \label{KMRSOM}
\omega \,=\,\bas\Lb 1 - \omega\Rb \Lb \frac{1}{\gamma}  + \frac{1}{1 - \gamma + \omega}\,+\,\underbrace{\Lb 2 \psi(1) - \psi\Lb 2 - \gamma\Rb -  \psi\Lb 1 + \gamma\Rb\Rb}_{\mbox{ high twist contributions}}\Rb
\eeq
One can see that $\gamma(\omega) \to 0$ when $\omega \to 1$, as follows
 from energy conservation.

The general solution to \eq{BFKL} takes the form:
 \beq \label{GENSOL}
 N\Lb  \xi ,  Y\Rb\,\,\,=\,\,\,\int^{\epsilon + i \infty}_{\epsilon - i \infty}\frac{d \gamma}{2\,\pi\,i} e^{\omega\Lb \bas,\gamma\Rb \,Y \,\,+\,\,\gamma\,\xi} \,\phi_{\rm in}\Lb \gamma\Rb
 \eeq
where $\gamma_{\rm in}$  can be found from the initial condition at
 $\eta = 0$. We suggest to take the initial condition in the form:

\beq \label{IC}
N\Lb \xi,  Y=0\Rb\,=\,\exp\Lb - B e^{-\,\h \xi}\Rb\,\,=\,\,
 \exp\Lb - B \frac{b^2}{r\,R}\Rb
\eeq

Taking the inverse Mellin transform, we obtain 
$\phi_{in}$ is equal to
\beq \label{PHIIN}
\phi_{in}\Lb \gamma\Rb\,\,=\,\,2\,B^{\,-\,2\,\gamma } \Gamma \Lb \,2\,\gamma \Rb
\eeq
One can see that  $\phi_{in}$ has a pole at $\gamma = 0$. To avoid 
this pole,  which occurs due to our
 simplifying the estimates, we modify  the initial conditions:
\beq \label{PHIIN1}
\phi_{in}\Lb \gamma\Rb\,\,=2\,\,\Lb B_1^{-2\,\gamma } \,-\, B_2^{-2\,\gamma }\Rb\,
\Gamma \Lb 2\,\gamma \Rb
\eeq
 \eq{PHIIN1} has no singularities at $\gamma = 0$, at any value of
 $B_1$ and $B_2$.

For large $Y$ and $\xi$ we can use the method of stepest descent in
 calculating the integral of \eq{GENSOL}.
  The equation for the saddle point
 ($\gamma = \gamma_{\rm SP}$)  is

\beq \label{SP}
\frac{d \omega\Lb \bas, \gamma\Rb}{d \gamma}\Big{|}_{\gamma \,=\,\gamma_{\rm SP}} \,\,=\,\,- \,\frac{\xi}{Y}
\eeq
For large $|\xi|$  ($|\xi|/Y\,\gg\,1)$ at $\gamma =
 \gamma_{\rm SP}$ $\frac{d \omega\Lb \bas, \gamma\Rb}{d \gamma}$
 should be large. All kernels, that we have discussed in
 \eq{CHI}\,-\,\eq{KMRSOM} are large at $\gamma \,\to\,1$ and, actually,
  accounting  for this singularity,  corresponds to the double log
 approximation(DLA) of perturbative QCD.


\section{ DLA for LO BFKL  equation}


 For the case of the leading order  BFKL equation at
 $\gamma\,\,\to\,\,1$,  $\omega_{\rm LO} \,=\,\frac{\bas}{1 -
 \gamma}$ and  \eq{SP} takes the form

\beq \label{SPLO}
\frac{\bas}{\Lb 1 - \gamma_{\rm SP}\Rb^2}\,Y\,\,= \,\,-\,\xi
\eeq
leading to $\bar{\gamma}_{\rm SP} \,\,\equiv\,\,1\,-\,\gamma_{\rm SP}
 \,=\,\,\sqrt{\frac{\bas\, Y}{|\xi|}}$.
 Plugging  this solution into \eq{GENSOL} we obtain that
 \beq \label{SOLLO}
 N\Lb Y, \xi \Rb\,\,\propto\,\,\phi_{in}\Lb \gamma_{\rm SP}\Rb \, \exp\Lb 2 \sqrt{\bas Y\,|\xi|} \,\,-\,\,|\xi|\Rb\,\,\to\,\,\phi_{in}\Lb \gamma_{\rm SP}\Rb\,\,\Lb \frac{b^4}{r^2\,R^2}\Rb^{ - 1 + 2 \bar{\gamma}_{\rm SP}}
 \eeq

Therefore, in the LO approximation we expect a power-like decrease
 of the scattering amplitude at large $b$, in accord
with the general discussion in Ref.\cite{KW1,KW2,KW3}.

The solution of \eq{SOLLO}  can be derived directly from \eq{KERLO}
 for the BFKL kernel. Indeed, DLA stems from $r' \gg r$ and the BFKL
 equation can be re-written  as follows
\beq \label{DLALO}
\frac{\partial}{\partial Y}N\Lb \xi  ,  Y  \Rb\,\,=\,\,\bas r^2\,\int_{r^2} \frac{d\,r'^2}{r'^4} N\Lb \xi' , Y\Rb\,
\eeq
 Substituting  $\tilde{N} =\,N\,e^{- \xi}$ and introducing a new
 variable $\tilde{\xi} = - \xi$ we see that \eq{DLALO} 
takes the following form:
\beq \label{DLALO1}
\frac{\partial^2 \tilde{N}\Lb \tilde{\xi}  ,  Y \Rb}{ \partial Y\,\partial \tilde{\xi}}\,\,=\,\,\bas\,\tilde{N}\Lb \tilde{\xi}  ,  Y \Rb
\eeq

Identifying $ N\Lb Y, \xi \Rb\,\propto\,e^{\gamma' \tilde{\xi}}$ 
we obtain the solution in the form  of \eq{SOLLO}.


\section{DLA for NLO BFKL}

\subsection{Generalities}
The large impact parameter behaviour of the scattering amplitude in
 the NLO BFKL equation is also determined by  the  values of $\gamma$,  which are close  to $\gamma = 1$ ( $\gamma \,\to\,1$).  The singular part of the general
 kernel in the NLO( see \eq{KERNLOR})  has the following form:
\beq \label{GANLOS}
\omega\,\,=\,\,\frac{\bas}{1 - \gamma + \omega};
\eeq
with the solution:
\beq \label{OMNLO}
\omega\Lb \gamma\Rb\,\,=\,\,\h\Lb-\Lb 1  - \gamma\Rb\,+\,\sqrt{4\,\bas \,+\,\Lb 1 - \gamma\Rb^2}\Rb
\eeq

Plugging \eq{OMNLO} into \eq{GENSOL}, we obtain the solution in the form:
\beq \label{SOLNLO}
 N\Lb \xi,  Y \Rb\,\,\,=\,\,\,\int^{\epsilon + i \infty}_{\epsilon 
- i \infty}\frac{d \gamma'}{2\,\pi\,i} e^{ \h\Lb- \gamma'\,+\,\sqrt{4\,\bas \,+\,\gamma'^2}\Rb\,Y \,\,+\,\,\gamma'\,\tilde{\xi}\,-\,\tilde{\xi}} \,\phi_{\rm in}\Lb \gamma', R\Rb
 \eeq
where we introduce $\gamma' = 1 - \gamma$ and $\tilde{\xi} \,=\,-\,\xi$.

 Note that \eq{KMRSOM} gives
\beq \label{OMNLOKMRS}
\omega\,\,=\,\,\frac{\bas}{1 - \gamma + \omega}\,\Lb 1\,-\,\omega\Rb; 
\eeq

All other terms in  \eq{KMRSOM} vanish at $\gamma = 1$.  Solving
 \eq{OMNLOKMRS} we obtain
\beq \label{OMNLOKMRS1}
\omega\,\,=\,\,\h\Lb-\Lb 1 \, -\, \gamma\,+\,\bas\Rb\,+\,\sqrt{4\,\bas \,+\,\Lb 1\, -\, \gamma\,+\,\bas\Rb^2}\Rb
\eeq

For $\omega$ of \eq{OMNLOKMRS1} the solution of \eq{SOLNLO} can
 be re-written as
\beq \label{SOLNLO1}
 N\Lb \xi, Y \Rb\,\,\,=\,\,\,\int^{\epsilon + i \infty}_{\epsilon - i \infty}\frac{d \gamma''}{2\,\pi\,i} e^{ \h\Lb- \gamma''\,+\,\sqrt{4\,\bas \,+\,\gamma''^2}\Rb\,Y \,\,+\,\,\gamma''\,\tilde{\xi}\,\,-\,\,\Lb 1 + \bas\Rb \tilde{\xi}} \,\phi_{\rm in}\Lb \gamma''\Rb
 \eeq
where $\gamma'' \,=\, \,\gamma' \,+\,\bas\,=\,1\,-\,\gamma\,+\,\bas$.


\subsection{DLA in coordinate representation}

Recently, a new approach to the NLO BFKL has been developed
 (see Ref.\cite{DIMST} and references therein) in which the most
 essential contributions were singled out and \eq{KERNLOR} has
 been resolved   with respect to $\omega$. The NLO BFKL is written
 in the coordinate representation in an elegant form with the
following kernel:
\beq \label{RESUMNLO}
K_{\rm rNLO}(x_{02},x_{12}; x_{10}) \,\, =\, \,K_{\rm LO} \Lb \vec{x}_{02}, \vec{x}_{12}; \vec{x}_{10}\Rb\, \left[\frac{x^{2}_{01}}{\min(x_{12}^{2}, x_{02}^{2})}\right]^{\pm \bas \,A_1} \frac{J_1(2\sqrt{\bas \rho^2})}{\sqrt{\bas \rho^2}}.
\eeq
where  the factor in square brackets leads to  the contribution of single
 collinear logarithms and factor $J_1(2\sqrt{\bas \rho^2}) / \sqrt{\bas
 \rho^2}$ resums double collinear logarithms to all orders. Parameter
 $A_1= 11/12$ and the sign in front of $A_1$ is positive, when
 $x^2_{01} < \min(r_{12}^2, r_{02}^2)$ and negative  otherwise. 
$J_1$  denotes the Bessel function (see formula {\bf 8.402} of
 Ref.\cite{RY}), $\rho\,\equiv\,\sqrt{L_{x_{02}, x_{01}}L_{x_{12}, x_{01}}}$
 and $L_{ x_{i 2}, x_{01}} \equiv \ln(x_{i 2}^2/x_{01}^2)$. The BFKL
 equation with the kernel of \eq{RESUMNLO} is solved in Ref.\cite{CCM}.
 It should be stressed,  that in the approach of Ref.\cite{DIMST}. the
 rapidity $Y$ should be replaced by  the target rapidity $\eta
 \,\,=\,\,Y \,-\,\ln\Lb \frac{R^2}{r^2}\Rb\,=\,\ln(1/x_{Bj})$
 for DIS scattering.

Finally, in the DLA  the BFKL equation in the re-summed NLO takes the form:

\beq \label{EQ}
\frac{d N\Lb r, b, Y\Rb}{d \eta}\,
\,=\,\,\bas \int_{r} \frac{d r'^2 \, r^2}{ r'^4}\,\,\frac{J_1\Lb 2  \sqrt{\bas \rho^2}\Rb}{\sqrt{\bas \rho^2}}\, N\Lb r', \vec{b} - \h(\vec{r'} - \vec{r})\Rb
\eeq

In \eq{EQ} we did not include  the factor
 $\Lb \frac{r^2}{r'^2}\Rb^{\bas A_1}$ for simplicity.
 It can be easily  be inserted  and has been taken into account
 in \eq{SOLNLO1}. The difference with \eq{OMNLOKMRS1} is
 that the argument $1 - \gamma + \bas$ should be replaced
 by $1 - \gamma +A_1\bas$.

Since in the DLA $\rho = \ln \kappa^2$ with $\kappa^2 = r^2/r'^2$ we
have  the following equations for the eigenvalues.

\begin{equation}\label{EIGENV}
\omega(\gamma,\bas) =\bas \displaystyle{ \int} \frac{dr'^2   r^2} { r'^4} \dfrac{J_{1}( \sqrt{2 \bas\rho^{2}}  )}{ \sqrt{\bas\rho^{2}}  } (\dfrac{r'}{r})^{2\gamma} \end{equation}
In the variable $\rho$ \eq{EIGENV} takes the form:
\begin{equation}\label{int2}
\displaystyle{ \int_{0}}^{1} dk^2  k^{-2\gamma} \dfrac{J_{1}( \sqrt{2 \bas } \ln  k^2  )}{ \sqrt{\bas} \ln  k^2  } =\displaystyle{ \int_{0}}^{\infty} d \rho  e^{-\rho\,(1-\gamma)} \dfrac{J_{1}( \sqrt{2 \bas } \rho  )}{ \sqrt{\bas} \rho  } \end{equation}
 From formulae {\bf 6.621(2)} of Ref.\cite{RY} and 
  {\bf 15.3.19} of Ref.\cite{ABST}
 
  \begin{equation}\label{int3}
\omega(\gamma,\bas) =  \dfrac{1}{2}(1-\gamma ) \Bigg(\sqrt{\frac{4\,\bas}{(1-\gamma ) ^2}+1}-1\Bigg) \,\xrightarrow{\bas \ll \,1\,\, \mbox{\tiny in LO BFKL}} \dfrac{ \bas }{(1-\gamma )} \end{equation}
Therefore, we see that  the  \eq{EQ}
 has the solution given by  \eq{SOLNLO}  for $\gamma' \,=\,1\,-\,\gamma$.

\subsection{Difficulties  present in  the method of steepest descent}


In the LO  to evaluate   the integral of \eq{GENSOL} we use 
the method
 of steepest descent. We now attempt to use it for the case of the NLO.

The  explicit equation for the saddle point  has the form (see 
\eq{SP}
 and \eq{int3}):

\beq \label{SP0}
\h\,\eta \left(\frac{1}{\sqrt{\frac{4 \bas }{\gamma'^2}+1}}\,-\,1\right) \,\,+\,\,\tilde{\xi}\,\,=\,\,0;~~~~
\h\,\eta \frac{1}{\sqrt{\frac{4 \bas }{\gamma'^2}+1}}\,\,=\,\,\h \,\eta\,\,-\,\,\tilde{\xi};
\eeq
From \eq{SP0} one can see that for $\h \eta \,>\,\tilde{\xi}$ the
 saddle point is  real, and we can obtain the reasonable asymptotic
 behaviour of the scattering amplitude. However, for $\tilde{\xi} \,>\,\h\,
 \eta$ the saddle point should be a complex number which, generally
 speaking,  leads to the oscillating behaviour, which contradict the
 unitarity constraint: $N\,>\,0$.

The solutions to \eq{SP0} are:
\bea \label{SPSOL}
\gamma_{\rm SP}\Lb \eta, \tilde{\xi}\Rb\,\,=\,\,\pm \frac{i\, \sqrt{\bas} \,|2\,\tilde{\xi} \,-\,\eta|}{\sqrt{\tilde{\xi} }\, \sqrt{\tilde{\xi} \,-\, \eta}}
 \left\{\begin{array}{l}\,\,\, \pm \,2\, i \sqrt{\bas} \Big( 1 + \frac{1}{8} \frac{\eta^2}{\tilde{\xi}^2}\Big) +O\left(\left(\frac{\eta}{\tilde{\xi}}\right)^{3}\right)\,\,\,\,\,\,\,\,\,\,\mbox{for}\,\,\,\tilde{\xi}\,\gg\,\,\eta \,\,\,\\ \\
\,\,\,\pm \Big( \sqrt{\frac{\bas\,\eta }{\tilde{\xi}}} \,\,-\,\,\frac{3}{2} \sqrt{\frac{\bas\,\tilde{\xi}}{ \eta}} \Big)\,\,+\,O\Lb \Lb \frac{\xi}{\eta}\Rb^{3/2}\Rb\,\,\,\,\,\mbox{for}\,\,\,\eta \,\gg\,\tilde{\xi},;  \end{array}
\right.
\eea

 From \eq{SPSOL} one can see, that we have two complex conjugate
 saddle points, which in general lead to an oscillating solution.
 Since $N$ is the imaginary part of the scattering amplitude, which
 is positive, we expect, that we will  have some difficulties with 
this method.
 
 As a  check to see whether we can  apply this method 
 successfully, we calculate $d^2
 \omega\Lb\gamma = \gamma_{\rm SP}\Rb /d \gamma^2$   and $d^3
 \omega\Lb\gamma = \gamma_{\rm SP}\Rb /d \gamma^3$, They have
 the following explicite forms:
 \beq \label{SPDER}
 \frac{d^2 \omega\Lb \gamma = \gamma_{\rm SP}\Rb}{ d \gamma^2 }\,\,=\,\,\frac{2 \bas }{\left(4\, \bas +\gamma^2_{\rm SP}\right)^{3/2}};~~~~~~~~~~~~~ \frac{d^3 \omega\Lb \gamma = \gamma_{\rm SP}\Rb}{ d \gamma^3 }\,\,=\,\,-\frac{6 \bas\,\gamma_{\rm SP} }{\left(4\, \bas +\gamma ^2_{\rm SP}\right)^{5/2}}\eeq
Plugging \eq{SP} in \eq{SPDER} we can see that 
\bea 
\tilde{\xi} \,\gg\,\eta  &~~~~~~& \eta \frac{d^2 \omega\Lb \gamma = \gamma_{\rm SP}\Rb}{ d \gamma^2 }\,\, \,\propto\,\,\,\frac{\tilde{\xi}^3}{\eta^2}; ~~~\eta \frac{d^2 \omega\Lb \gamma = \gamma_{\rm SP}\Rb}{ d \gamma^2 }\,\,\,\propto\,\,\,\frac{\tilde{\xi}^5}{\eta^4};
\label{SPDER1}\\
  \eta  \,\gg\,\tilde{\xi}  &~~~& \eta \frac{d^2 \omega\Lb \gamma = \gamma_{\rm SP}\Rb}{ d \gamma^2 }\,\, \,\propto\,\,\,\frac{\tilde{\xi}^{3/2}}{\eta^{1/2}}; ~~~\eta \frac{d^2 \omega\Lb \gamma = \gamma_{\rm SP}\Rb}{ d \gamma^2 } \,\,\,\propto\,\,\,\frac{\tilde{\xi}^2}{\eta};
\label{SPDER2}
\eea
From \eq{SPDER1} we  see that taking the Gaussian integral
 $\exp\Bigg(\h \eta \frac{d^2 \omega\Lb \gamma = \gamma_{\rm SP}\Rb}{
 d \gamma^2 } \Lb \gamma \,-\gamma_{\rm SP}\Rb^2\Bigg)$
 we obtain  the
 typical $\Lb \gamma \,-\gamma_{\rm SP}\Rb\,\propto 1/\sqrt{\h \eta
 \frac{d^2 \omega\Lb \gamma = \gamma_{\rm SP}\Rb}{ d \gamma^2 }}
 \,\propto \eta/\tilde{\xi}^{3/2}$.
 Inserting this estimate into
 $\exp\Bigg(\frac{1}{6}\eta \frac{d^3 \omega\Lb \gamma = \gamma_{\rm SP}\Rb}{
 d \gamma^3 } \Lb \gamma \,-\gamma_{\rm SP}\Rb^3\Bigg)$
 we see that this contribution is large ( proportional to
 $ \exp\Lb \sqrt{\tilde{\xi}}/\eta\Rb$).  This shows
 that we cannot use the method of steepest decent, at least
 for $\tilde{\xi} \gg \eta$.
 It should be noted that \eq{SPDER2} leads to a small contribution
 of the term of the order $ \Lb \gamma \,-\gamma_{\rm SP}\Rb^3$,
 in accord with the 
method of steepest descent. It is instructive to note that these
 conclusions  are in accord with the values of
 $\gamma_{\rm SP}$, which is pure imaginary at $\tilde{\xi}
 \,\gg\,\eta$ and real for $ \eta \,\gg\,\tilde{\xi}$.
\subsection{Expansion in series}


First, we re-write \eq{SOLNLO} in  a slightly different form as
\beq \label{SOL2}
\tilde{N}\Lb \xi,  Y\Rb\,\,=\,\,\int^{\epsilon + i \infty}_{\epsilon - i \infty}\frac{d \gamma'}{2 \pi i}e^{ \h \sqrt{ 4 \bas \,+\,\gamma'^2}\,Y\, \,+\,\,\gamma'\Lb \tilde{\xi} \,\,-\,\,\h\,Y\Rb} \,\phi_{in}(\gamma')
\eeq
\eq{SOL2} we expand in the following way
\beq \label{SOL3}
\tilde{N}\Lb \tilde{\xi},  Y\Rb\,\,=\,\,\int_{C_1}\frac{d \gamma'}{2
 \pi i}e^{ \gamma'\Lb \tilde{\xi} \,\,-\,\,\h\,Y\Rb} \,\phi_{in}(\gamma')
\,\sum^\infty_{n=0} \frac{\Lb \h \sqrt{ 4 \bas \,+\,\gamma'^2}\,Y\Rb^n}{n!}
\eeq

\begin{figure}
\begin{tabular}{ccc}
      \includegraphics[width=7cm]{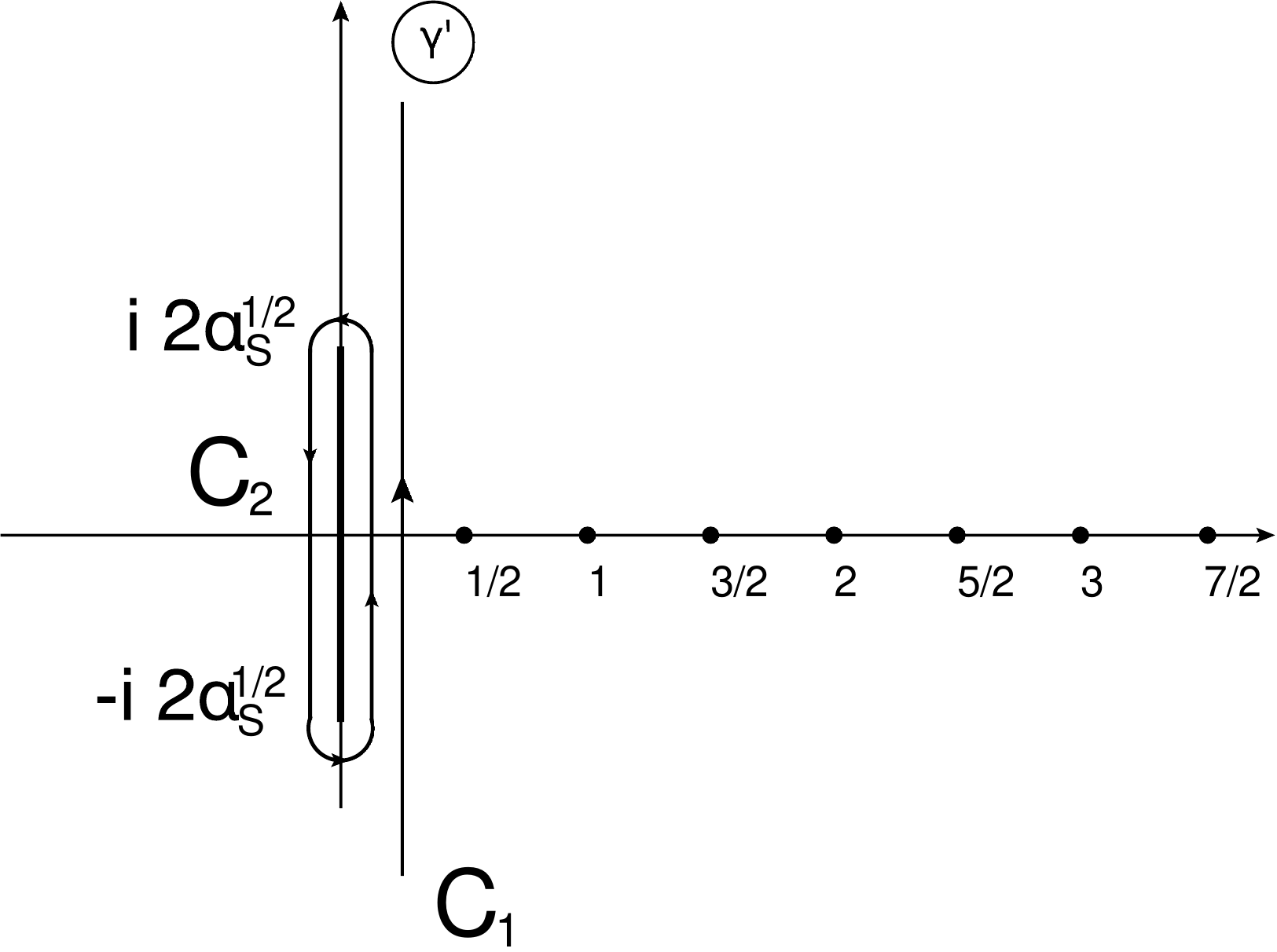}  &~~~~~~~~&\includegraphics[width=7cm]{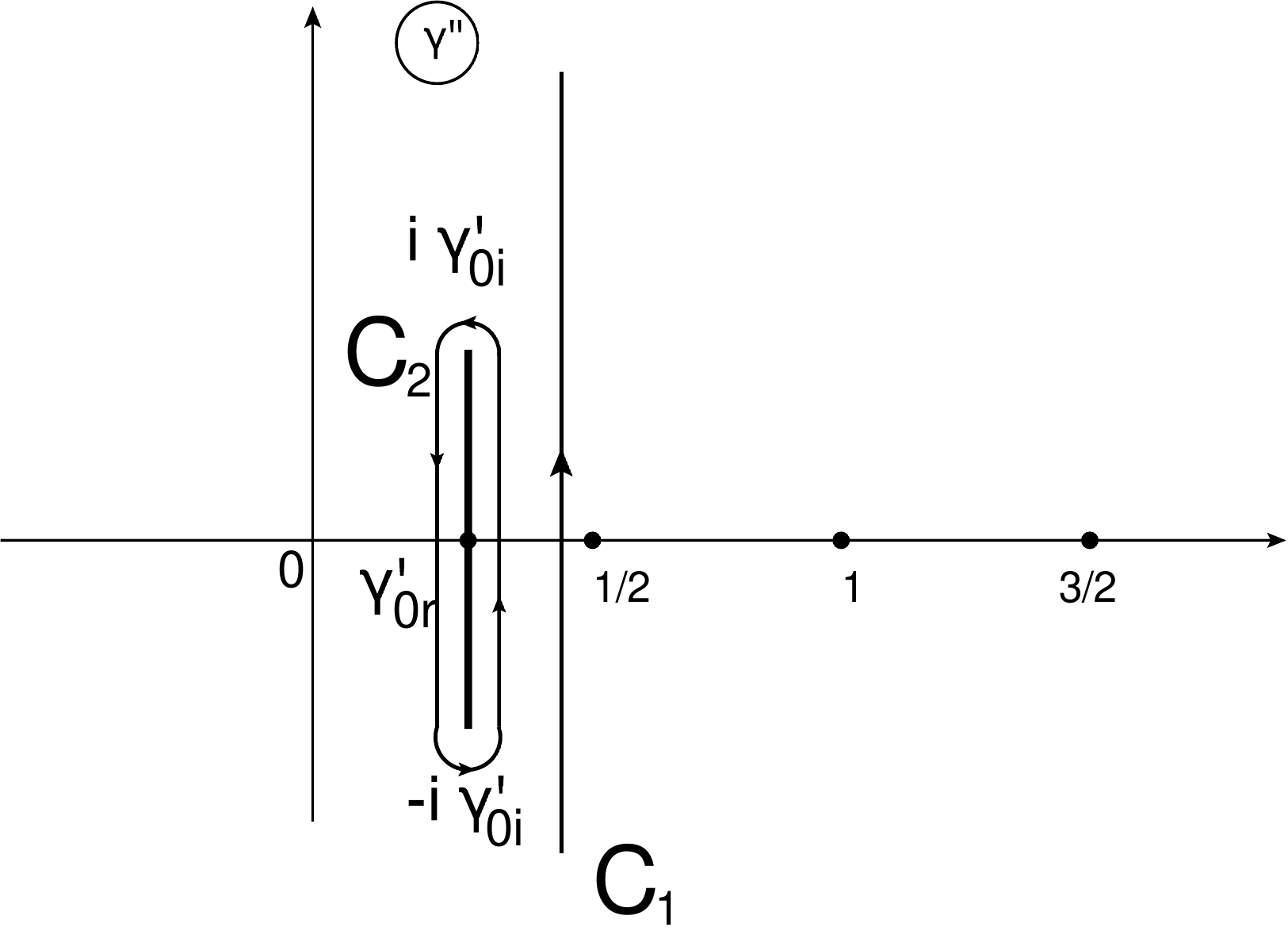} \\
      \fig{cont}-a & & \fig{cont}-b\\
      \end{tabular}
         \caption{The  $\gamma$ - plane: the contours  of integrations 
over $\gamma'$ in \eq{SOLNLO} and in \eq{SOL3} (\fig{cont}-a); and the 
contours of integrations in general case (\fig{cont}-b).
         }
        
\label{cont}
\end{figure}

In \fig{cont}-a   we plot the contour $C_1$ for the integration
 in \eq{SOL3}. Each term has singularities 
 in the right semi-plane,
 at points $n/2$, from $\phi_{in}(\gamma)$ (see \eq{PHIIN}) and also
 every term with  even  $n$   has 
singularities: the branch point
 from $ -\, i \,2\, \sqrt{\bas}$ to $ i \,2\, \sqrt{\bas}$.  For
 $\xi - \h Y  \,>\,0$ we can move the contour $C_1$ to the left
 and integrate each term with the contour $C_2$. Note, that for
 $\xi - \h Y  \,<\,0$  we can close the contour on the singularities
 of the initial conditions, or  make an analytical continuation of
 the scattering amplitude from the region $\xi - \h Y  \,>\,0$. 
   For large $\xi (Y - \xi)$, we can use the method of
 steepest descend to obtain the answer in this kinematic region.

 Hence the solution can be written in the form:

\beq \label{SOL4}
\tilde{N}\Lb \tilde{\xi},  Y\Rb\,\,=\,\,\frac{1}{ 2\,\pi}\int_{C_2}  \!\!\!\!d \gamma' \,e^{i\, \gamma'\Lb \tilde{\xi} \,\,-\,\,\h\,Y\Rb} \,\phi_{in}(i\,\gamma')\,\sum^\infty_{n=0} \frac{\Lb \h \sqrt{ 4 \bas \,-\,\gamma'^2}\,\,\,Y\Rb^{2 n\,+\,1}}{( 2 n \,+\,1)!}\eeq

For small values of $\bas$ we can safely replace $\gamma'$ by $\gamma'=0$
 in $\phi_{in}\Lb i \gamma'\Rb$ and take the integral, using formula 
 {\bf 3.771(8)} of Ref. \cite{RY}:
\bea \label{INT}
&&\int^{2 \sqrt{\bas}}_{- 2 \sqrt{\bas}}  \!\!\!\!d \gamma' \,e^{i\, \gamma'\Lb \tilde{\xi }\,\,-\,\,\h\,Y\Rb} \, \frac{\Lb \h \sqrt{ 4 \bas \,-\,\gamma'^2}\,\,\,Y\Rb^{2 n\,+\,1}}{( 2 n \,+\,1)!}\,\,= \\
&&\,\,=\,\phi_{in}\Lb 0\Rb \,\,\displaystyle{2\,\sqrt{\pi}(2\,\as)^{n+1} Y^{2n+1}\dfrac{\Gamma\left(n+\frac{3}{2}  \right)}{\Gamma(2n+2)}  J_{n+1}(\sqrt{\as}\vert 2\, \tilde{\xi}-Y\vert  )(\sqrt{\as} \vert 2\,\tilde{\xi}-Y\vert )^{-(n+1)}   }\nn\\
&&\,\,=\,\,\phi_{in}\Lb 0\Rb\,\,\displaystyle{   \dfrac{2 \sqrt{\as} \pi Y}{\vert  2\,\tilde{\xi}-Y\vert} \dfrac{1}{n!} J_{n+1}(\sqrt{\as}\vert 2\,\tilde{\xi}-Y \vert  ) \left(2\,\dfrac{(\sqrt{\as} Y/2)^{2}}{\sqrt{\as}\vert 2\, \tilde{\xi}-Y\vert   }\right)^{n}  }\nn
\eea
where we use the  duplication formula of the Gamma function(see formula
 {\bf 8.335(1)} of Ref.\cite{RY}):
$\Gamma\Lb 2(n + 1)\Rb\,=\,\Lb 2^{2 n + 1}/\sqrt{\pi}\Rb\Gamma\Lb n + 1
  \Rb\,\Gamma\Lb n+3/2\Rb$.

\begin{figure}[ht]
  \begin{tabular}{ccc}
      \includegraphics[width=7cm]{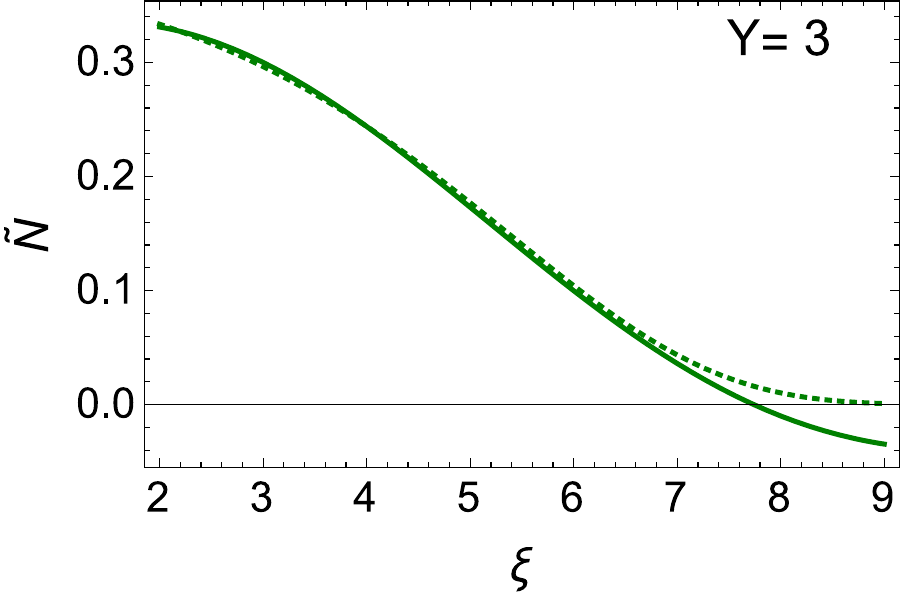} &  ~~~~~~~ & \includegraphics[width=7cm]{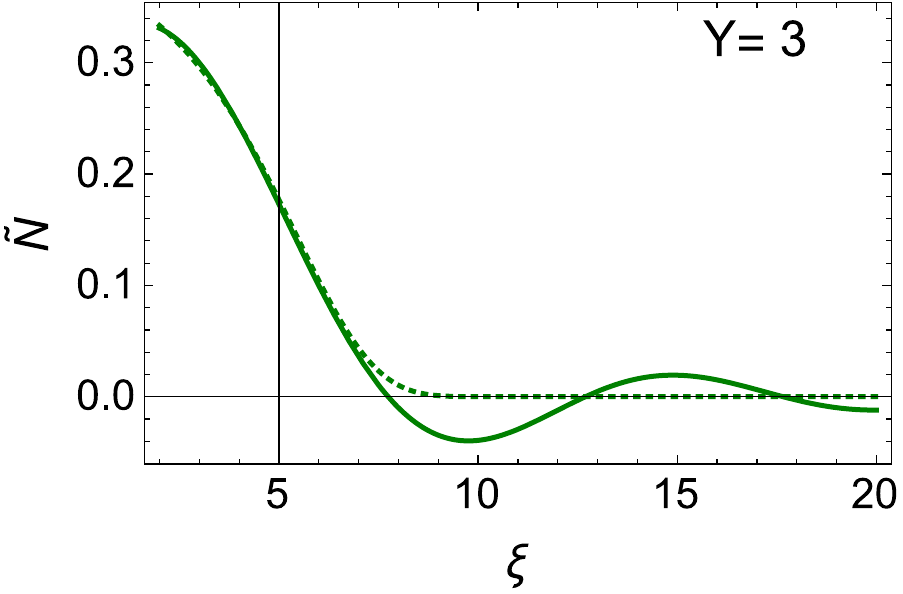} \\
      \fig{sol}-a&  & \fig{sol}-b\\
       \includegraphics[width=7cm]{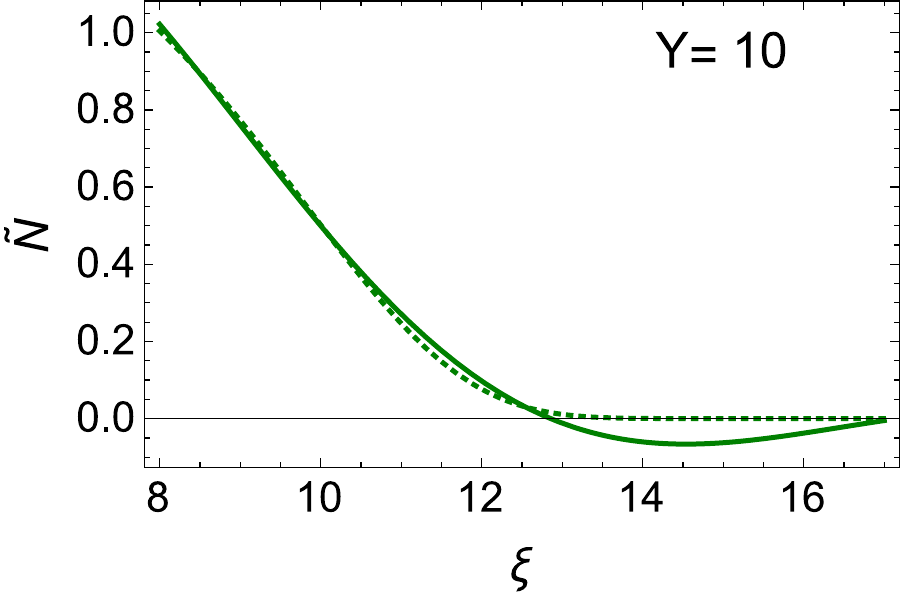} &  ~~~~~~~ & \includegraphics[width=7cm]{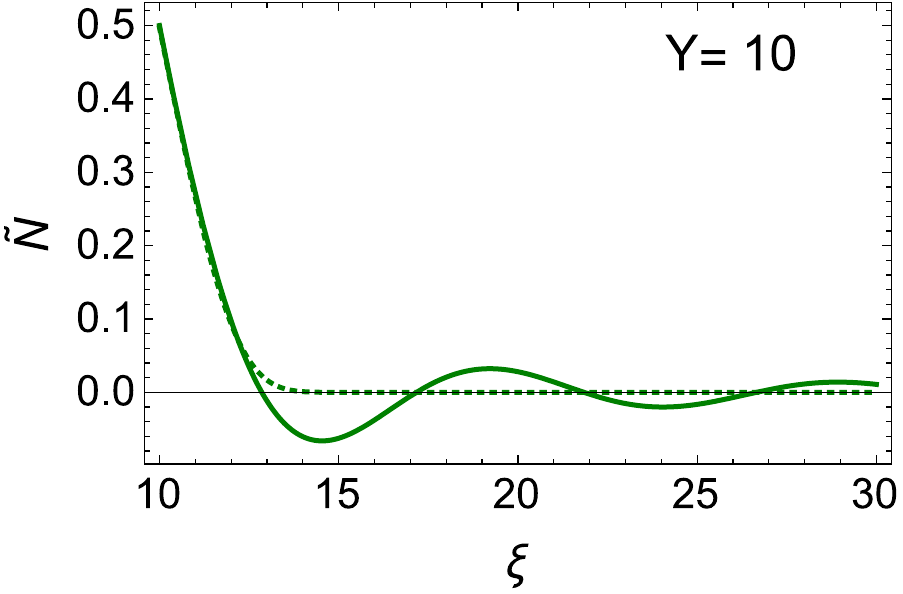} \\
      \fig{sol}-c&  & \fig{sol}-d\\      
      \end{tabular}
      \caption{Solution  $\tilde{N}\Lb Y,\xi\Rb $ of \eq{ANS2}
 versus $\xi$ (solid lines). The dotted lines are the fit $\tilde{N}
 \propto \exp\Lb - \mu^2 b^2 \Rb$ with $\mu^2 = 0.07/(r R)$ for
 \fig{sol}-a and \fig{sol}-b and $\mu^2 \,=\,0.007/(r R)$ for
 \fig{sol}-c and \fig{sol}-d. In all estimates $\bas=0.2$.}
\label{sol}
   \end{figure}

Plugging \eq{INT} into \eq{SOL3} we obtain 

\beq \label{ANS}
N(\tilde{\xi},Y)\,\,\,=\,\,\, e^{- \tilde{\xi}}\,\,\phi_{in}\Lb 0 \Rb\,\,\dfrac{\sqrt{\bas } Y}{\vert 2\,\tilde{\xi}-Y\vert} \displaystyle{  \sum_{n=0}^{\infty}  \dfrac{\tau^{n} }{n!} J_{n+1}(\sqrt{\bas}\vert \,2\,\tilde{\xi}-Y \vert  ) }
\eeq
with $\tau\,\,=\,\,2\,(\sqrt{\as}\,\h Y )^{2}/ (\sqrt{\bas}\vert
 2\,\tilde{\xi}-Y \vert )$. For $\vert\tilde{\xi}-Y \vert^{2}>
 Y^{2}/2$ the series term can be summed  using formula
 {\bf 5.7.6.1} in Ref.\cite{PR}. Hence,  we obtain the
 explicit form of the solution

 \beq\label{ANS2}
N\Lb\tilde{\xi},Y\Rb\,\,\,=\,\,\,e^{- \tilde{\xi}}\,\,\phi_{in}\Lb 0\Rb\,\, \displaystyle{  \bas  Y  \dfrac{J_{1}\Lb 2\,\sqrt{\bas\, \tilde{\xi} \,\Lb \tilde{\xi} - Y\Rb}\Rb}{ \sqrt{\bas\, \tilde{\xi} \,\Lb \tilde{\xi} - Y\Rb}  }}
\eeq
In \fig{sol} we plot $\tilde{N}\Lb\tilde{\xi},Y\Rb$. 
 For the LO BFKL equation this function increases with
 $\tilde{\xi}$.
From \fig{sol} one can see that (i) at large $b$ the
 solution decreases as the power of $b$; (ii) in the
 limited 
range of $\xi$ we can parameterize this decrease as
 $\tilde{N} \propto \exp\Lb - \mu^2 b^2 \Rb$ with
$\mu^2 ={\rm Const}/(r R)$  for sufficiently small
 values of ${\rm Const}$;  and (iii) at large $b$ we
 have oscillating behaviour, which is in contradiction
  to $\tilde{N} \,>\,0$, that follows from the unitarity constraints.

All these features can be seen from the asymptotic behaviour of
 \eq{ANS2} at large $\xi \,\gg\,Y$.
 One can see that the scattering amplitude
 \beq 
 \label{ASBE}
 N(\tilde{\xi},Y)\,\,\,\propto \,\,\,e^{- \tilde{\xi}}\frac{ \cos\Lb \frac{\pi}{4} - 2 \,\sqrt{\bas}\,\tilde{\xi}\,\Rb}{\Lb 2 \sqrt{\bas}\,\tilde{\xi}\Rb^{3/2}}\,\,
 \leq\,\,\frac{r^2\,R^2}{b^4}
 \eeq
 Therefore, we have  power-like bahaviour of the scattering amplitude at
 large $b$, which leads to the violation of the Froissart theorem 
\cite{KW1,KW2,KW3}.

For $\tilde{\xi} \,<\,Y$   the solution takes the form
\beq \label{ANS3}
N\Lb\tilde{\xi},Y\Rb\,\,\,=\,\,e^{- \tilde{\xi}}\,\,\phi_{in}\Lb 0\Rb\,\bas  Y\,\Lb  \displaystyle{  \dfrac{I_{1}\Lb 2\,\sqrt{\bas\, \tilde{\xi} \,\Lb Y -  \tilde{\xi} \Rb }\Rb}{ \sqrt{\bas\, \tilde{\xi} \,\Lb Y -  \tilde{\xi} \Rb}  } }\Rb\eeq

Therefore, the general solution can be written as
\bea \label{ANS4}
&&N\Lb\tilde{\xi},Y\Rb\,\,\,=\\
&&\,\,\,e^{- \tilde{\xi}}\,\,\phi_{in}\Lb 0\Rb\,\bas  Y\,\Lb  \displaystyle{  \dfrac{I_{1}\Lb 2\,\sqrt{\bas\, \tilde{\xi} \,\Lb Y -  \tilde{\xi} \Rb }\Rb}{ \sqrt{\bas\, \tilde{\xi} \,\Lb Y -  \tilde{\xi} \Rb}  } } \,\Theta\Lb Y \,- \,\tilde{\xi}\Rb \,\,+\,\, \displaystyle{  \dfrac{J_{1}\Lb 2\,\sqrt{\bas\, \tilde{\xi} \,\Lb \tilde{\xi} - Y\Rb }\Rb}{ \sqrt{\bas\, \tilde{\xi} \,\Lb \tilde{\xi} - Y\Rb}  } } \,\Theta\Lb  \tilde{\xi}\,-\,Y\Rb\Rb\nn
\eea

This solution, is similar to the solution of the BFKL equation with
  time ordering (see Eq.(3.35) in Ref.\cite{DIMST}), if we replace
 $\tilde{\xi}$  by $\rho\,=\,\ln\Lb R^2/r^2\Rb$. We cannot
 claim that $\tilde{\xi}\,>\, Y$ corresponds to the unphysical
 kinematic region due to the  time ordering,  since the BFKL kernel
 does not depend on  impact
parameters.

\subsection{Numerical estimates}

\begin{figure}[h]
      \includegraphics[width=9cm]{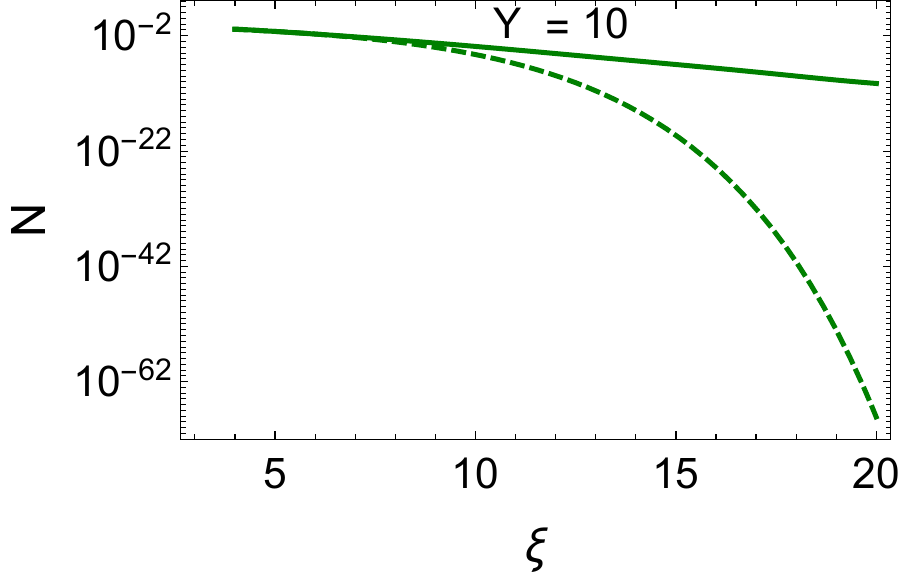}  
    \caption{Numerical estimates for  $N$, which comes from  \eq{SOL6}
 fixing $B_1= \h$ and $B_2 = 1$ in \eq{PHIIN1},$\bas = 0.2$ and $Y = 10$
 (solid line) . The dotted line is $N\, \propto\, \exp\Lb - 1.06\,
 e^{\frac{1}{4} \xi}\Rb$.}
\label{sol1}
\end{figure}

 Summing over $n$ in \eq{SOL4} we can re-write the solution in the form:
 \beq \label{SOL6}
 \tilde{N}\Lb \xi,  Y\Rb\,\,=\,\,\frac{1}{2\,\pi}\int^{2 \sqrt{\bas}}_{- 2 \sqrt{\bas}} d \gamma' \Bigg(  e^{ \h \sqrt{ 4 \bas \ - \,\gamma'^2}\,Y}\,\,-\,\,e^{ -\h \sqrt{ 4 \bas \ - \,\gamma'^2}\,Y} \Bigg)\,\,e^{i\,\gamma'\Lb \xi \,\,-\,\,\h\,Y\Rb} \,\phi_{in}(i\,\gamma')
\eeq 
  In \fig{sol1} we plot $N$, which comes from the numerical calculation
 for \eq{SOL6}, choosing $B_1 = \h$ and $B_2 = 1$ in \eq{PHIIN1}, 
 taking $\bas = 0.2$ and fixing $Y=10$. The logarithmic plot in
 this figure shows, first, that at large $b$ we have the power-like
 decrease, as we have discussed, and, second, 
   that we can reproduce the solution which decreases as
 $e^{ - 1.06 b/\sqrt{r\,R}}$ in the region of $\xi = 4 - 10$.
  It should be stressed that such fast decrease  cannot be
 achieved in the LO BFKL, for which, $\tilde{N}$ increases at
 large $b$. We will discuss this in detail in the conclusions below.
 It is interesting to note that the slope $1.06/\sqrt{r \,R}$ is close
 to one, that has been found in Ref.\cite{CCM} for $r = R = 10 \,GeV^{-1}$.

\begin{figure}[ht]
  \begin{tabular}{ccc}
      \includegraphics[width=7cm]{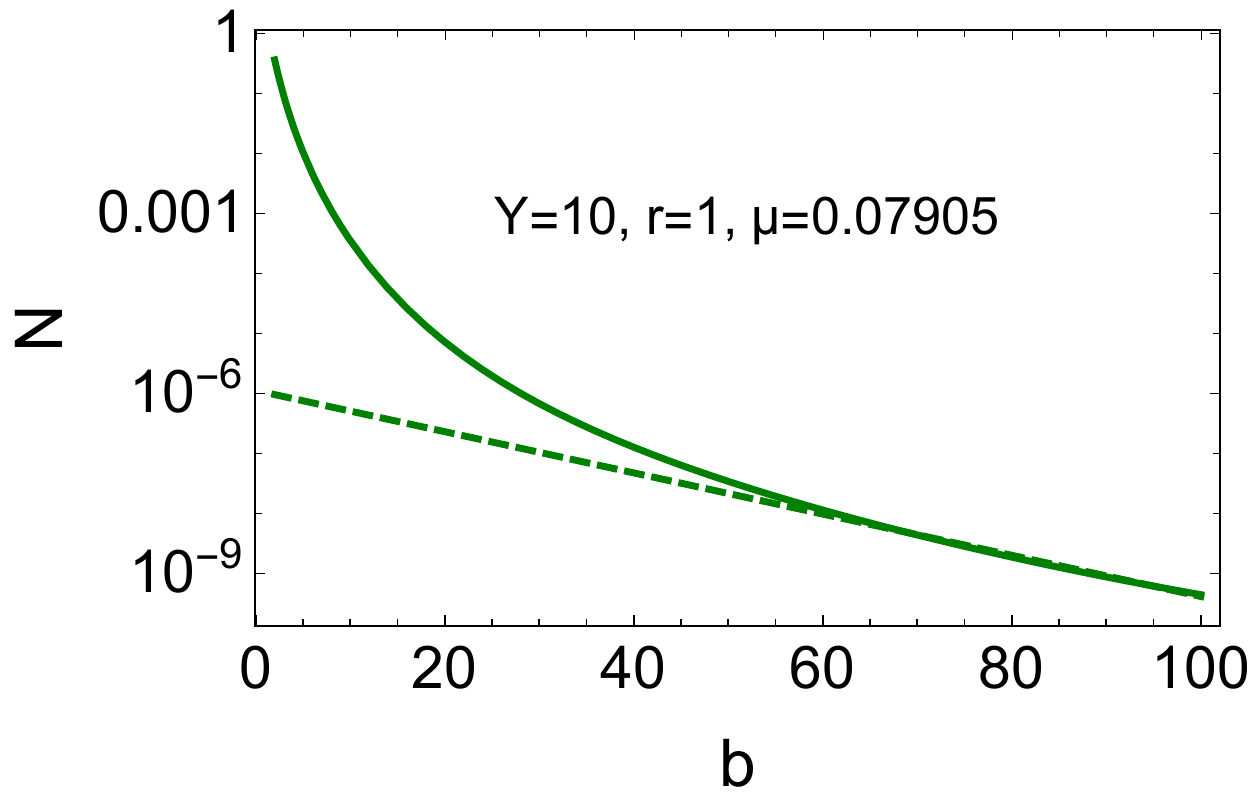} &  ~~~~~~~ & \includegraphics[width=7cm]{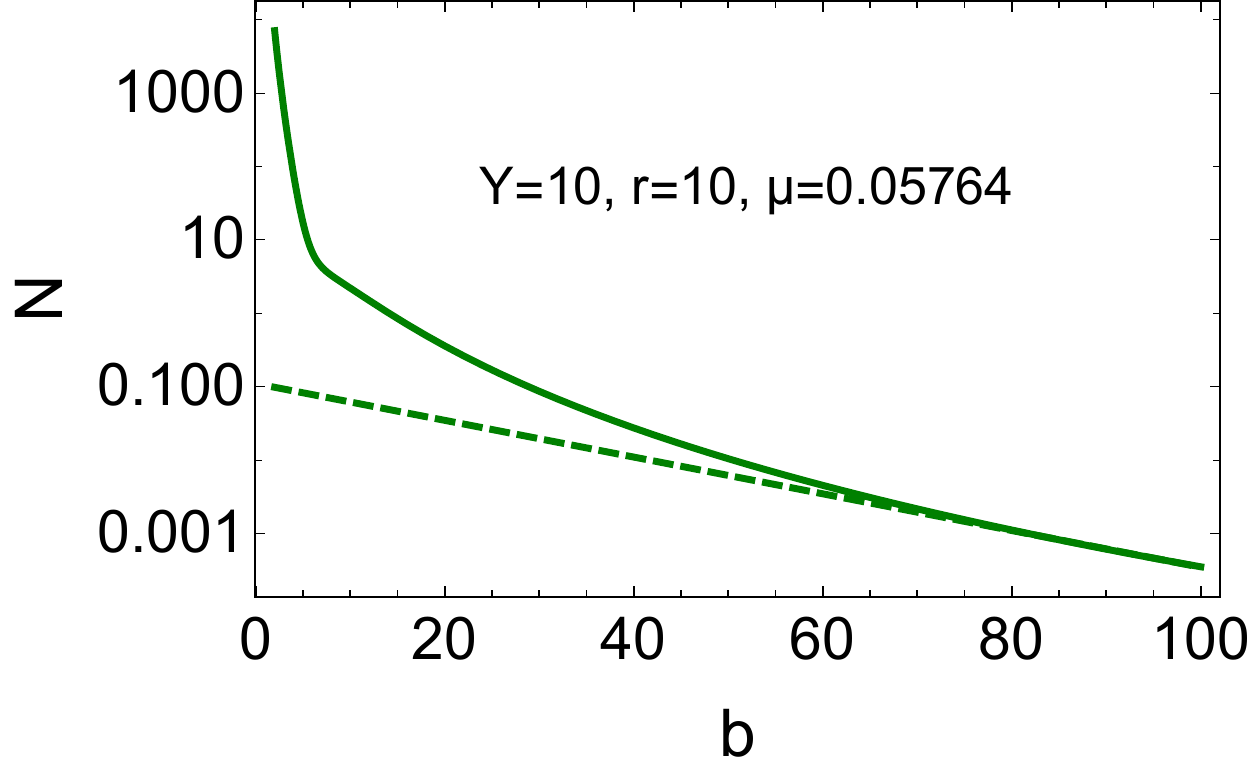} \\
      \fig{num}-a&  & \fig{num}-b\\
       \includegraphics[width=7cm]{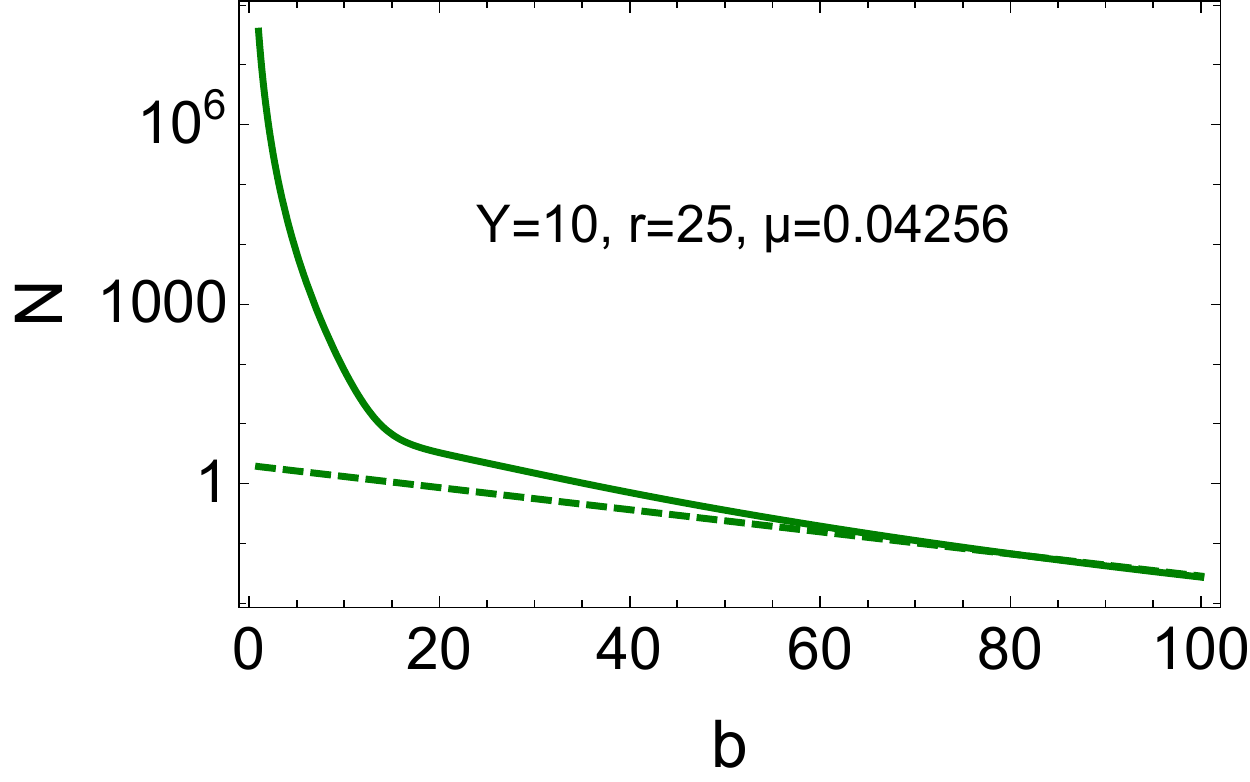} &  ~~~~~~~ & \includegraphics[width=7cm]{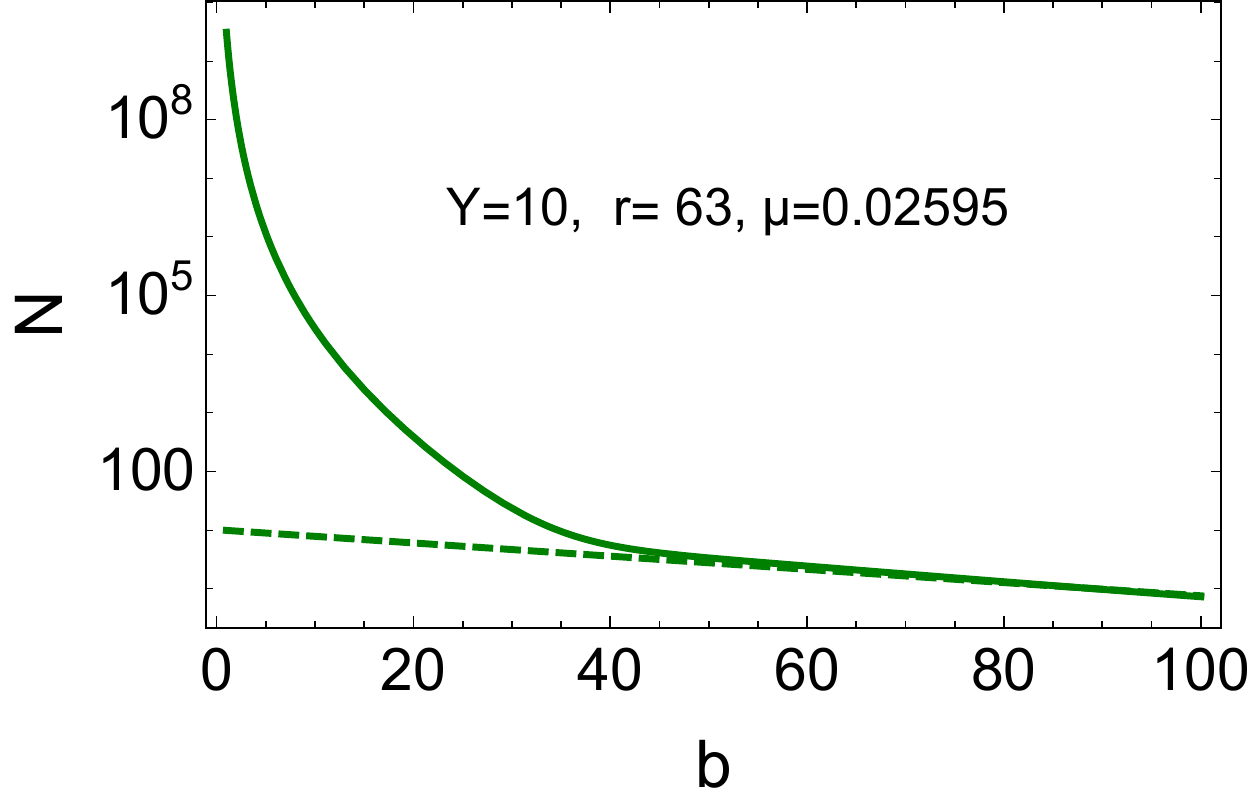} \\
      \fig{num}-c&  & \fig{num }-d\\      
      \end{tabular}
      \caption{Numerical estimates of \eq{SOL6} for different values of
 $r = R$ at $Y = \ln(1/x)$ = 10 (solid lines). The dotted lines denote $N \propto \exp\Lb - \mu b\Rb$. The slope $\mu$ is in $GeV$, while $r$ 
and
 $b$ are in $GeV^{-1}$. For all estimates $\bas=0.2$.}
\label{num}
   \end{figure}

 Tables I and II as well as \fig{num} show that the slope 
$\mu$($N\, \propto\, e^{- \mu\,b}
 $) depend on the values of $Y$ and on the initial conditions.
 One can see that the range of $b$ in which we can trust the
 exponential parameterization  also depends on the values of
 $r$  and $Y$, reproducing the main pattern of the solution
 given in Ref.\cite{CCM}.
 
  \begin{table}[h]
\begin{minipage}{8cm}{
\centering
\hspace{-3cm}
\begin{tabular}{|c|l|l|}
\hline
r ($GeV^{-1}$) &   Y=10&  Y=3\\  \hline
1  & 0.079 & 0.394  \\
 10& 0.058 & 0.092  \\
 25 & 0.043 & 0.082  \\
63 & 0.026 & 0.038\\\hline
\end{tabular}
\caption{ Numerical value for the \\slope $\mu$ in $GeV$ versus $Y =
 \ln(1/x)$ }
\label{Table1}
}
\end{minipage}
\begin{minipage}{8cm}
{\begin{tabular}{|c|c|c|c|}
\hline 
r ($GeV^{=1}$)  & $B_1$=1/2,$B_2$ =1 & $B_1$=1/2,$B_2$ =2 & $B_1$=1/3,$B_2$ =1
\\ \hline
1  & 0.079 & 0.067  & 0.076 \\
 10& 0.058 & 0.053  & 0.057 \\
 25 & 0.043 & 0.036 & 0.039  \\
63 & 0.026 & 0.023& 0.022\\
\hline
\end{tabular}
\caption{ Numerical value for the  slope $\mu$ in $GeV$ for $Y =10$ and 
 for different values of $B_1$  and $ B_2$.  }
\label{Table2}}
\end{minipage}
\end{table}
 
\section{Beyond DLA} 
 
 In this section we   modify the solution 
  taking into account  more complicated expressions for the
 eigenvalues than \eq{OMNLO} and \eq{OMNLOKMRS1}. We consider 
\eq{KMRSOM} and the eigenvalues $\omega\Lb \bas, \gamma\Rb$ take
 the form
 \bea \label{BDLA1}
&&\omega\Lb \bas, \gamma\Rb\,\,=\\
&&\,\,\underbrace{ \frac{-\bas +\gamma -1+\bas  \gamma  \Phi (\gamma )}{2 (\alpha  \Phi (\gamma )+1)}}_{\omega'\Lb \bas, \gamma\Rb} \, +\,\underbrace{\frac{\sqrt{(\bas  (-\gamma ) \Phi (\gamma )+\bas -\gamma +1)^2-4 (\bas  \Phi (\gamma )+1) (\bas  \gamma  \Phi (\gamma )-\bas  \Phi (\gamma ) - \bas)}}{2 (\alpha  \Phi (\gamma )+1)} }_{\omega''\Lb \bas, \gamma\Rb}\nn
\eea
with
\beq \label{PHIO}
 \Phi (\gamma )\,\,=\,\,\frac{1}{\gamma}\,  +\, 2 \psi(1) - \psi\Lb 2 - \gamma\Rb -  \psi\Lb 1 + \gamma\Rb
 \eeq
 
\begin{figure}
      \includegraphics[width=9cm]{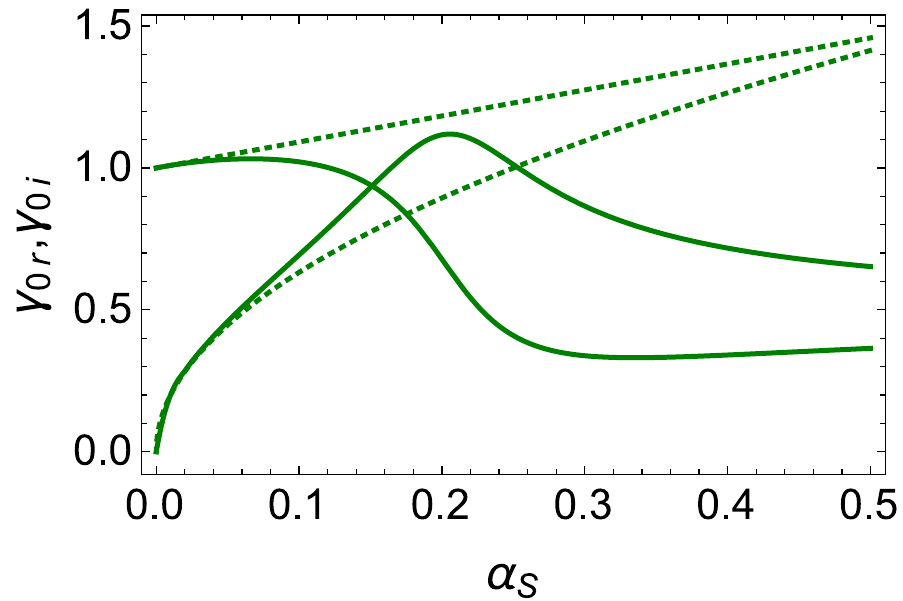}  
    \caption{$\gamma_{0 r}$(solid line)  and  $\gamma_{0 i} $( dotted line)
 versus $\bas$. The roots for the 
    DLA approach (see \eq{OMNLOKMRS1})  are shown by the dotted lines. In
 this case $\gamma_{0 r} = 1  + \bas$ and $\gamma_{0 i}\,=\,2 \bas^{1/2}$}
\label{root}
\end{figure}

  %


The singularities of  $\omega\Lb \bas, \gamma\Rb $ are related to
 the poles of $\Phi\Lb \gamma\Rb$, the zeroes of $ 1 + \bas \Phi\Lb
 \gamma\Rb=0$ and $\omega"\Lb \bas, \gamma\Rb $ has a branch point
 when the radicand is 
equal to zero. Near the zero of the radicand
 $ \omega"\Lb \bas, \gamma\Rb $  takes the form
\beq \label{OMPP}
\omega"\Lb \bas, \gamma\Rb\,\,=\,\,A\Lb \bas\Rb \sqrt{ \Lb \gamma
 \,\,-\,\,\gamma_{0 r} \Rb^2 \,+\,\gamma^2_{0 i}}
\eeq
\eq{OMPP} has two complex  roots: $ \gamma\,\,=\,\,\gamma_{0 r}\,\,\pm\,
\,i\,\gamma_{0 i}$. In \fig{root} we plot $\gamma_{0 r}$ and  $\gamma_{0
 i} $ as functions of $\bas$. From this figure shows that for very small
 values of $\bas$ our solution coincides with the DLA approximation.
 However, for $\bas \,>\,0.05$ the values of $\gamma_{0r}$ and
 $\gamma_{0i}$ differ considerably from their DLA values, approaching
 their maxima at large $\bas$.

The contours of the integration over $\gamma' = 1 - \gamma$ are shown
 in \fig{cont}-b. The integration over contour $C_2$ in \fig{cont}-b 
can be written in the form
\bea \label{BDLA2}
\tilde{N}\Lb \xi, Y\Rb\,\,&=&\,\,\frac{1}{2\,\pi}\int^{\gamma_{0i}}_{- \gamma_{0i}}\,d \gamma''  \exp\Lb \omega'\Lb \bas, \gamma = \gamma_{0r}\Rb\, Y \,+\,i\,\frac{d \omega'\Lb \bas, \gamma = \gamma_{0r}\Rb}{d \gamma} \gamma'' \,Y\,+\,i \gamma''\, \xi\Rb\nn\\
& \times&\,\,
\Lb \exp\Lb A\Lb \bas\Rb\,Y\, \sqrt{  \gamma^2_{0i} \,\,-\,\,\gamma''^2 }\Rb \,\,-\,\,\exp\Lb- A\Lb \bas\Rb\,Y\,\sqrt{  \gamma^2_{0i} \,\,-\,\,\gamma''^2 }\Rb\Rb
\eea
Introducing the new notation: $\omega_0 \Lb \bas\Rb\,=\,\omega'\Lb \bas,
 \gamma = \gamma_{0r}\Rb$
and $B\Lb \bas\Rb\,=\,\frac{d \omega'\Lb \bas, \gamma =
 \gamma_{0r}\Rb}{d \gamma} $
we can  evaluate the integral of \eq{BDLA2} using the same procedure,
 as we have discussed in section IV-D, or  using   formula {\bf 3.711}
 of Ref.\cite{RY}   continuing it analytically for imaginary $A$. 

Finally,

\beq \label{BDLA3}
\tilde{N}\Lb \xi, Y\Rb\,\,=\,\,Y\,\gamma_{0i}\,A\Lb \bas\Rb  e^{ \omega_0\Lb \bas\Rb Y}\frac{J_1\Lb  \gamma_{0,i}\sqrt{\Lb \xi + B\Lb \bas\Rb Y\Rb^2 \,\,-\,\,A^2\Lb \bas\Rb Y^2}\Rb}{\sqrt{\Lb \xi + B\Lb \bas\Rb Y\Rb^2 \,\,-\,\,A^2\Lb \bas\Rb Y^2}}
\eeq
\begin{figure}
      \includegraphics[width=9cm]{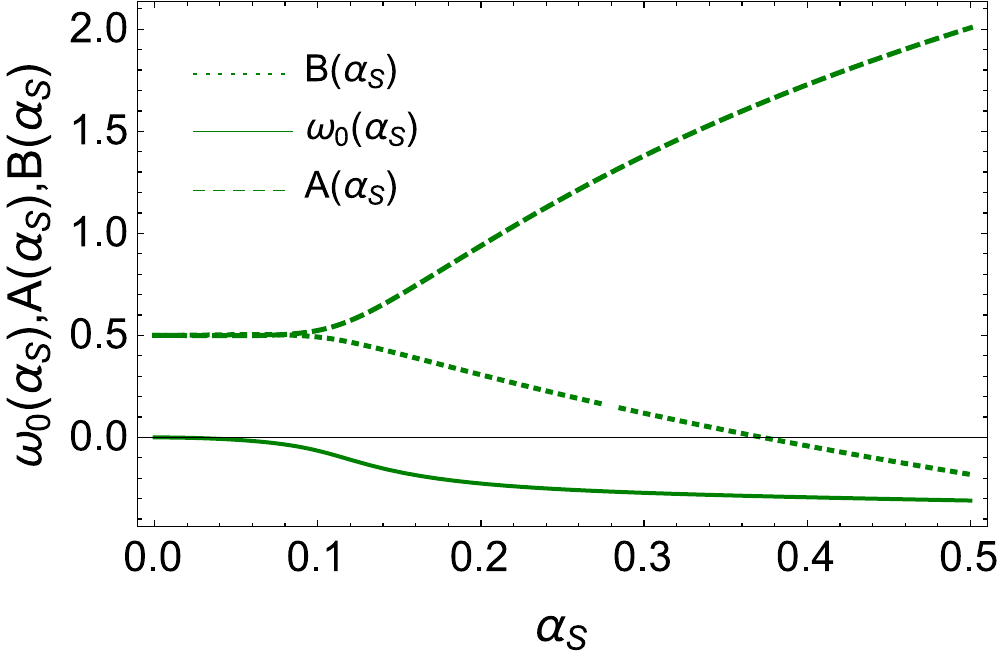}  
    \caption{$\omega_0\Lb \bas\Rb,A\Lb \bas\Rb,B\Lb \bas\Rb$ versus $\bas$.}
\label{omab}
\end{figure}

The $\bas$ dependence of all parameters in   \eq{BDLA3} are shown in \fig{omab}.

\begin{figure}[ht]
\begin{tabular}{c c c}
 \includegraphics[width=6cm]{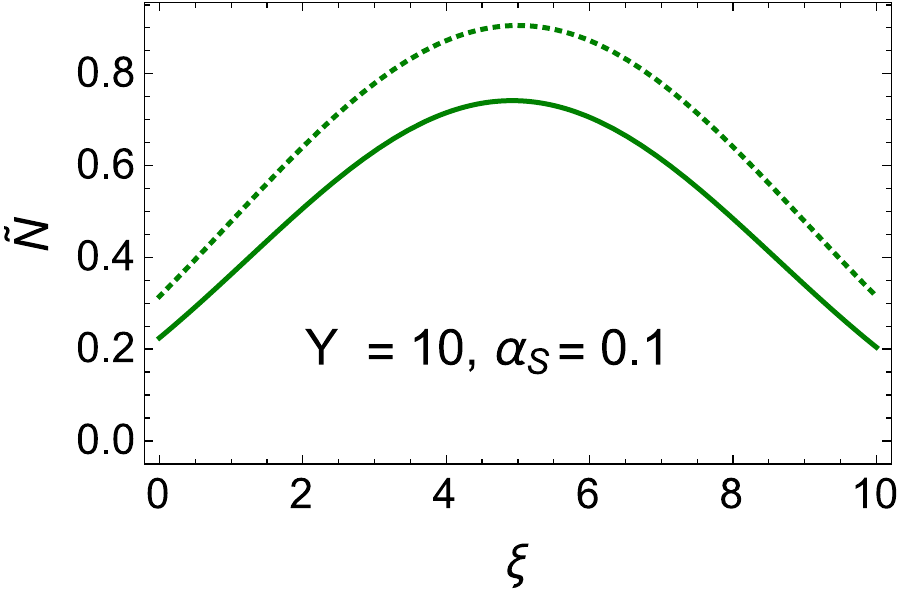}&~~~~& \includegraphics[width=6cm]{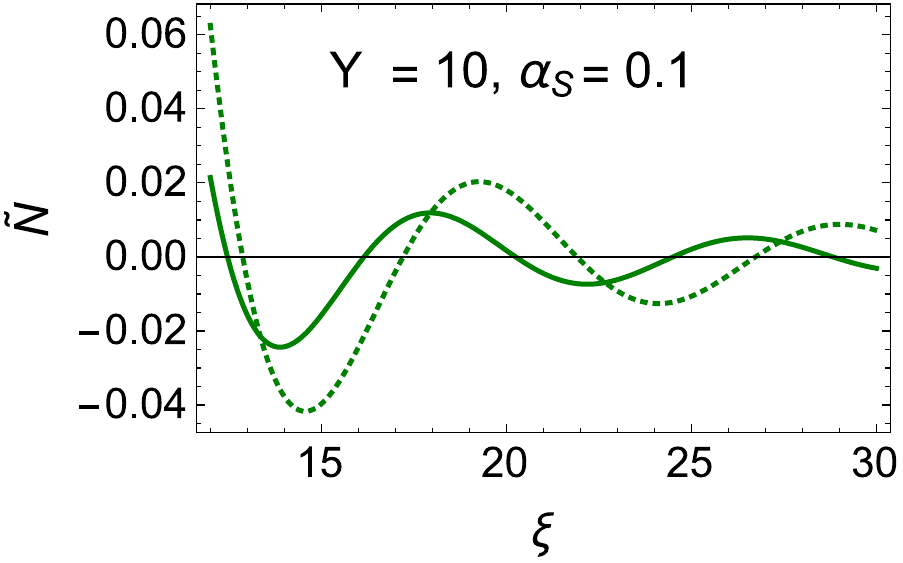}\\
 \fig{fit}-a &  & \fig{fit}-b\\
  \includegraphics[width=6cm]{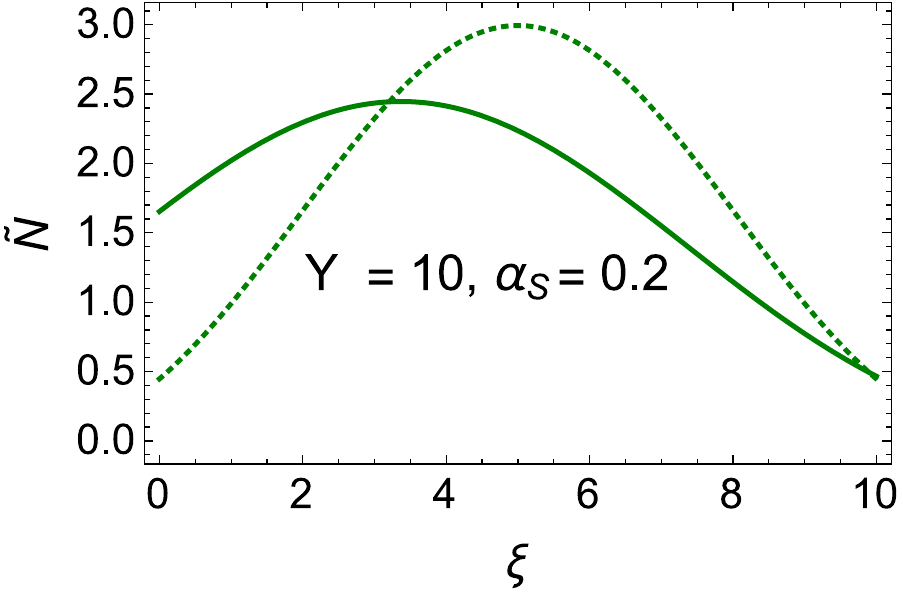}&~~~~& \includegraphics[width=6cm]{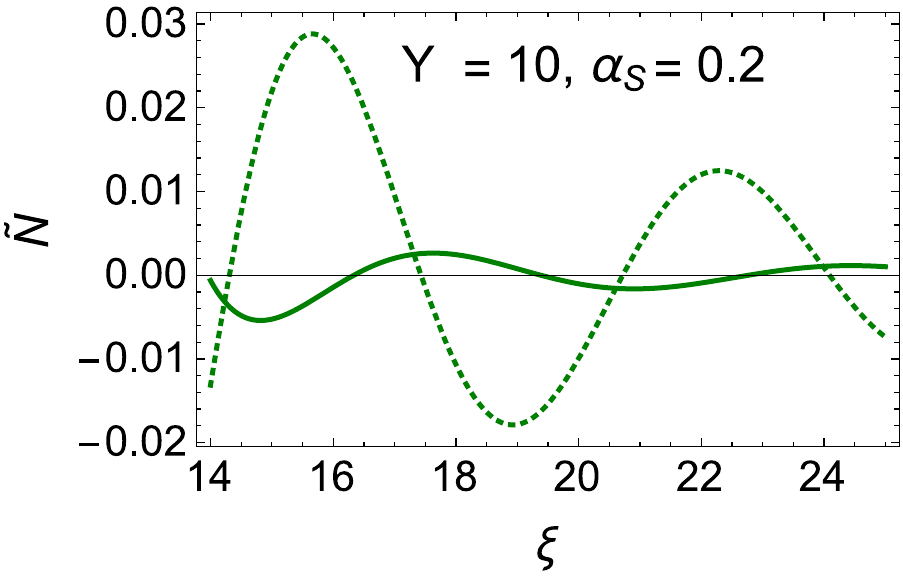}\\
 \fig{fit}-c &  & \fig{fit}-d\\ 
 \includegraphics[width=6cm]{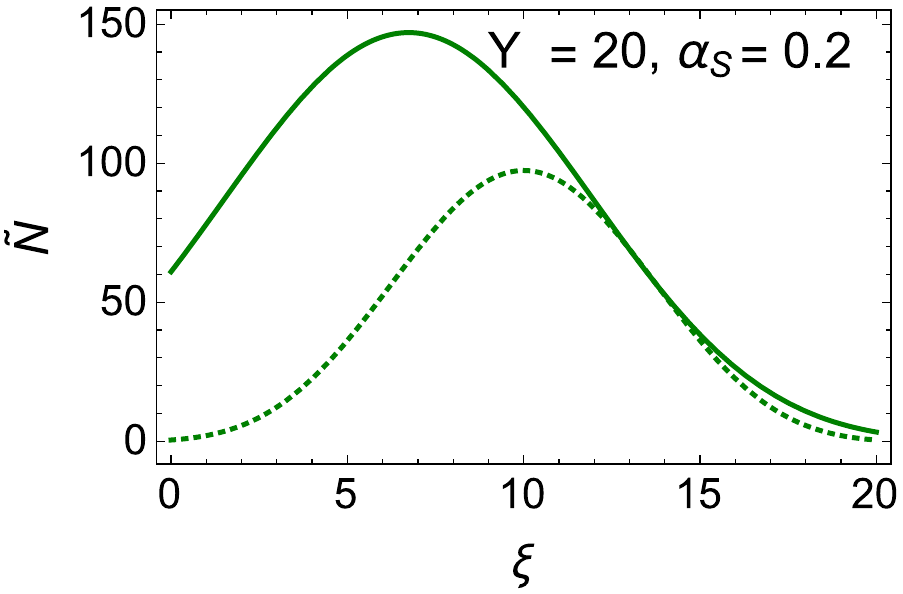}& ~~~~~&\includegraphics[width=6cm]{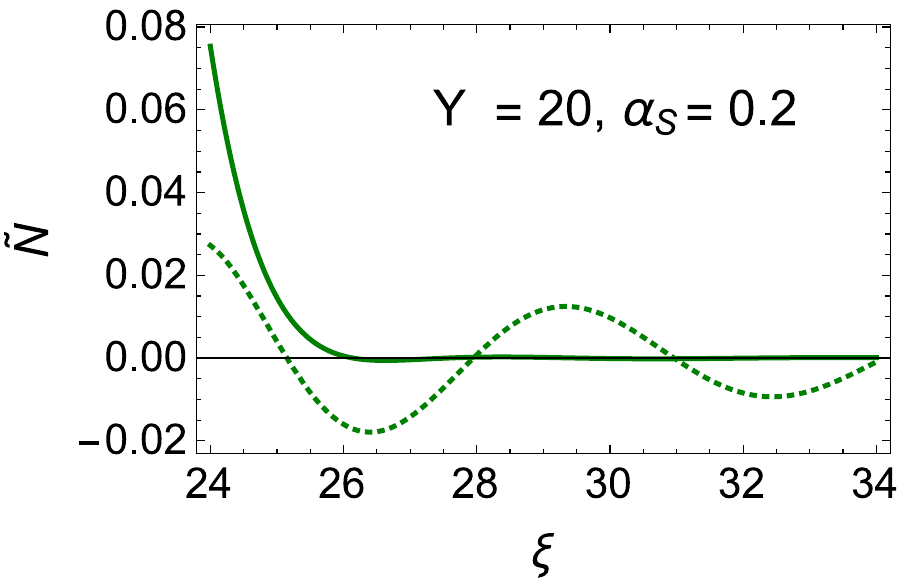}\\
 \fig{fit}-e &  &\fig{fit}-f\\ 
 \end{tabular}
    \protect\caption{ Solutions of \eq{BDLA3} (solid lines) and 
solution of \eq{ANS4} (dotted lines) in different kinematic regions. 
 }
\label{fit}
   \end{figure}

In \fig{fit} we give several examples of the behaviour of $\tilde{N}$
 in different kinematic regions. One can see that in spite of  numerical
 differences, the claim that $\xi \,<\,Y$ give the main contribution 
 is correct.

\section{Conclusions}

  In this paper, we show that the NLO corrections  do not change the 
power-like decrease of the scattering amplitude  at large impact
 parameter and, therefore, they cannot   resolve the contradiction
 with the unitarity\cite{KW1,KW2,KW3}.  On the other hand, in a
 limited range of $b$, the NLO corrections lead to 
 a fast decrease  of the scattering amplitude with $b$,
 which can
 be parameterized as $N\, \propto\,\exp\Lb -\,\mu^2\,b^2\Rb$ with
 $\mu^2 \propto 1/r^2$, in accord with the numerical estimates in
 Ref.\cite{CCM}.
 
We demonstrate  that the NLO correction leads to an oscillating
 behaviour of the scattering amplitude as function of $b$. Such
 oscillations contradict  the unitarity constraints, as 
   $N$, being the imaginary part of the scattering amplitude,
  should be positive ($N\,>\,0$). 
 
\begin{figure}
      \includegraphics[width=9cm]{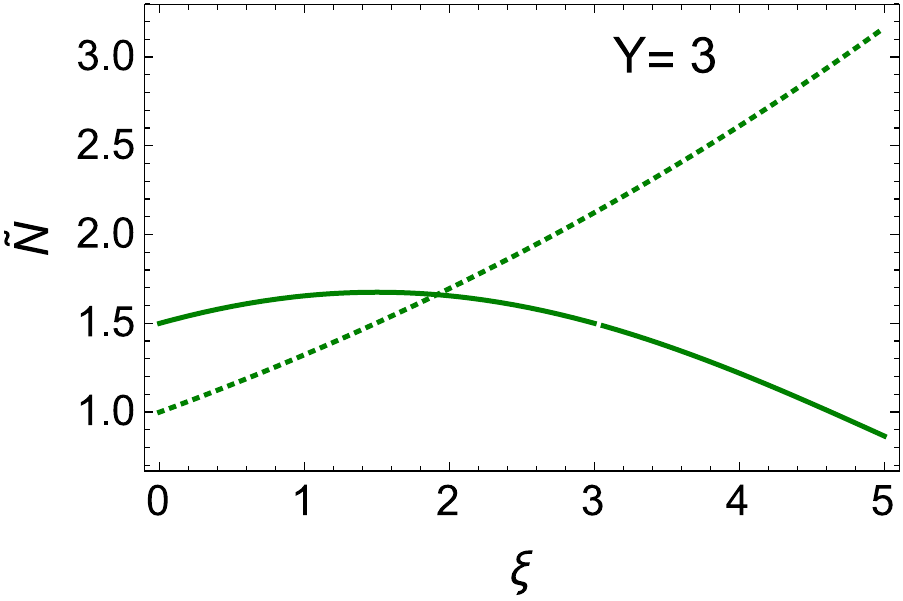}  
    \caption{Comparison of the NLO calculation(solid line) with
 the LO estimates (dotted line).$\bas$ = 0.2. Y = 3.}
\label{sol2}
\end{figure}

 However, from the more practical point of view, the NLO estimates
 give the faster decrease of the scattering amplitude as a function
 of $b$ (see \fig{sol2}) and could be  useful in the description
 of the experimental data (see Ref.\cite{CCM}).

 In a sense, we showed that the scattering amplitude is negligibly
  small at $\tilde{\xi} \,>\,Y$( $b^2 > r^2\,\exp\Lb \h \,\eta \Rb$).
  The violation of the Froissart theorem stems from the smaller values
 of $\xi$. Indeed, for $\xi \,<\,Y$ the scattering amplitude is proportional
 to $ N \,\propto\,\exp\Lb  2 \sqrt{\bas\,\xi\,\Lb Y\,-\,\xi\Rb }\,-\,\xi\Rb$
(see \eq{ANS4})   and $N \ll\,1 $ for $\xi\,\geq\,4 \bas Y/(1 + 4\,\bas)$.
 Choosing $\xi_0 = \ln\Lb b^2_0/r^2\Rb$ we see that
 \beq \label{FTH}
\sigma \,=\,2\,\int d^2 b  N\,\,\leq\,\,2 \pi \int^{b^2_0}d\, b^2\,\,\sim\,\,b^2_0\,=\,r^2 e^{ 4 \bas \eta/(1 + 4\,\bas)}\,\gg\,Y^2
\eeq 
 Therefore, the range of $b^2 \, <\, r^2\,e^{ 4 \bas \eta/(1 + 4\,\bas)} $
 turns out to be wide enough to violate the Froissart
 theorem\cite{KW1,KW2,KW3}. Hence, the resumed NLO kernel cannot heal the
 problem of violation of the Froissart theorem and has an additional
 defect of the oscillating behaviour at $\xi > Y$, which is in  
 contradiction to the unitarity constraints, which lead to  $N>0$.

 We believe that we need to introduce non-perturbative corrections with an
 additional dimensional scale  to the BFKL kernel, and  that their 
influence will
 be much more important than  that of the NLO BFKL kernel that we 
have discussed here.

  \section{Acknowledgements}
   We thank our colleagues at Tel Aviv university and UTFSM for
 the  discussions. The special thanks go to Asher Gotsman  his encouraging 
  support. This research was supported  by 
   Proyecto Basal FB 0821(Chile),  Fondecyt (Chile) grants  
 1180118 and 1191434.

  \end{document}